\documentclass[10pt]{article}

\usepackage[margin=25mm]{geometry}

\usepackage{url,hyperref,lineno,microtype,caption,subcaption}
\usepackage{microtype}
\usepackage{amsmath, amsfonts, amssymb}
\usepackage{bm}
\usepackage[table,xcdraw]{xcolor}
\usepackage{dsfont}
\usepackage{palatino}
\usepackage{algorithm, algpseudocode}

\usepackage{tabularx}
\usepackage{makecell}
\usepackage{booktabs}
\usepackage{multirow}
\usepackage{enumitem}
\usepackage{hhline}

\usepackage{mathtools}
\mathtoolsset{showonlyrefs}

\usepackage{setspace}
\newcommand{\sM}{\mathcal{M}}
\newcommand{\sN}{\mathcal{N}}

\newcommand{\sS}{\mathcal{S}}

\newcommand{\sR}{\mathcal{R}}%reflection operator
\newcommand{\lr}[2][{}]{\langle #2 \rangle}
\newcommand{\dfn}{\stackrel{\text{def}}{=}}
\newcommand{\walpha}{\widetilde{\alpha}}
\newcommand{\wsN}{\widetilde\sN}

\newcommand{\rvline}{\hspace*{-\arraycolsep}\vline\hspace*{-\arraycolsep}}

\title{The mass-conversion method: a hybrid technique for simulating well-mixed chemical reaction networks}
\date{}
\author{Joshua C. Kynaston\footnote{Department of Mathematical Sciences, University of Bath} \footnote{Author to whom correspondence should be addressed. Email: josh.kynaston@bath.edu} \and Christian A. Yates\footnotemark[1] \and Anna Hekkink \and Chris Guiver\thanks{School of Engineering \& The Built Environment, Edinburgh Napier University} }

\begin{document}
\maketitle

\begin{abstract}
    There exist several methods for simulating biological and physical systems as represented by chemical reaction networks. Systems with low numbers of particles are frequently modelled as discrete-state Markov jump processes and are typically simulated via a stochastic simulation algorithm (SSA). An SSA, while accurate, is often unsuitable for systems with large numbers of individuals, and can become prohibitively expensive with increasing reaction frequency. Large systems are often modelled deterministically using ordinary differential equations, sacrificing accuracy and stochasticity for computational efficiency and analytical tractability. In this paper, we present a novel hybrid technique for the accurate and efficient simulation of large chemical reaction networks. This technique, which we name the mass-conversion method, couples a discrete-state Markov jump process to a system of ordinary differential equations by simulating a reaction network using both techniques simultaneously. Individual molecules in the network are represented by exactly one regime at any given time, and may switch their governing regime depending on particle density. In this manner, we model high copy-number species using the cheaper continuum method and low copy-number species using the more expensive, discrete-state stochastic method to preserve the impact of stochastic fluctuations at low copy number. The motivation, as with similar methods, is to retain the advantages while mitigating the shortfalls of each method. We demonstrate the performance and accuracy of our method for several test problems that exhibit varying degrees of inter-connectivity and complexity by comparing averaged trajectories obtained from both our method and from exact stochastic simulation.
\end{abstract}

\section{Introduction} %!%--1 Introduction--%%%%%%%%%%%%%%%%%%%%%%%%%%%%%%%%%%%%%%%%%%%%%%%%%%%%%%%%%%%%%%%%%%%%%%%%%%%%
  A chemical reaction network (CRN) is a representation of a reacting (bio)chemical system of several species interacting via some number of reaction channels. CRNs, such as those found in biological systems, are often represented by continuous time, discrete-state Markov processes. This modelling regime is appropriate when the described system has a small number of interacting particles and provides an exact description of reaction dynamics under appropriate assumptions; specifically, that the inter-event times between the `firing' of reaction channels are independent and exponentially distributed. Such Markov processes are most often simulated via a stochastic simulation algorithm (SSA), the prototypical example of which is the Gillespie direct method~\cite{gillespie_exact_1977}. Several improvements to the Gillespie direct method have been proposed for reaction networks with particular structural characteristics. For example, the next reaction method~\cite{gibson_efficient_2000} and the optimised direct method~\cite{cao_efficient_2004} are exact and efficient SSAs for systems with a large number of loosely-coupled reaction channels. Further extensions also exist, such as the modified next reaction method~\cite{anderson_modified_2007}, that facilitate the simulation of systems with time-dependent reaction rates.

  For any reaction network, and under mild differentiability assumptions, one can derive a partial differential equation called the \textit{chemical master equation} (CME) that describes the time-evolution of the probability density of the system existing in any given state~\cite{gillespie_rigorous_1992}. The CME, as a single equation that encapsulates all stochastic information of a system, is neither solvable analytically nor practicable to solve numerically in all but the most straightforward of systems. Rather, the practical utility of the CME lies in the ease with which one can derive time-evolution equations for the raw moments of the system. These moment equations take the form of a system of ordinary differential equations (ODEs) that govern the moments of each constituent species. In cases where the CRN contains reactions of at least second-order, these moment equations do not form a closed system; in particular, the equations governing the $n^\text{th}$ moments will, in general, depend on the $(n+1)^\text{th}$ or higher-order moments. These systems are not solvable analytically. As such, one generally applies a so-called `moment-closure' that closes the system of moment equations at a given order by making explicit assumptions about the relationships between lower- and higher-order moments. Commmon moment-closures (or, simply, closures) include the mean-field closure, wherein all moments above the first are set to zero, and the Poisson closure, where diagonal cumulants are assumed equal to their corresponding mean and all mixed cumulants are set to zero~\cite{schnoerr_comparison_2015}.

  In general, determining the most appropriate closure assumptions for a given system can be challenging and higher-order closures often yield systems of moment equations that can be difficult to solve; as such, straightforward closures like the mean-field see the widest application. In the case of the mean-field closure, the resulting system of mean-field ODEs provides an approximate, continuous, and deterministic description of the time evolution of the mean of the underlying Markov process, and can be solved either analytically or numerically.

  The primary downside of SSAs is that they may become computationally intractable for large systems of interacting particles. Even for systems with favourable network structures, large systems can quickly become infeasible to simulate exactly. This is contrasted with deterministic modelling techniques that sacrifice accuracy in exchange for computational efficiency where, notably, the efficiency of numerical simulation methods (i.e., those for ODEs and PDEs) is typically independent of copy number. The various advantages and disadvantages of each modelling regime discussed have motivated the development of so-called hybrid methods that combine regimes to leverage their advantages and mitigate their limitations (see e.g.~\cite{smith_spatially_2018}). Several general hybrid approaches have been developed to tackle these issues. One such approach is to model certain species under a continuous regime (such as an ODE or SDE) and others under a discrete regime (via a SSA). Typically, this extension of the system is accomplished by categorising reactions as either being `fast' or `slow', applying a continuous representation to the former and using a discrete method for the latter. Cao, Gillespie, and Petzold~\cite{cao_slow_2005} pioneered this technique in the development of the `slow-scale SSA', a method for simulating dynamically stiff chemical reaction networks. Their method separates reactions and reactant species into fast and slow categories in a manner that allows for only the slow-scale reactions and species to be simulated stochastically, subject to certain stability criteria of the fast system. The fast-slow paradigm was also applied by Cotter,~\textit{et al.}~\cite{cotter_constrained_2011} for simulating chemical reaction networks that can be extended into fast and slow `variables', which may be reactant species or combinations thereof. They define a `conditional stochastic simulation algorithm' that can draw sample values of fast variables conditioned on the values of the slow variables. These samples are then used to approximate the drift and diffusion terms in a Fokker-Planck equation that describes the overall state of the system.

  In this paper we detail the development of a novel hybrid simulation technique for well-mixed CRNs; that is, systems of interacting (bio)chemical species distributed homogeneously within a reactor vessel of fixed volume. As discussed, continuum methods are advantageous when copy numbers are high and the effects of stochasticity can be safely assumed to be small. Discrete methods, on the other hand, are best applied in low copy number systems and where stochasticity is a critical driver of the dynamics. It is this fundamental tension between computational efficiency and model accuracy that our method seeks to address. Where other, similar methods aim to subdivide species and/or the reactions between them into categories based on reaction rates, we take a simpler approach that is instead based on particle density. Our objective is to create a method that is simple to implement, computationally efficient, accurate, and flexbile enough to handle not only reaction networks with fast/slow reactions, but also more uniform reaction networks where no such fast/slow distinctions can be leveraged.

  Our method, which we term the \textit{mass conversion method} (MCM), consists of a system of ODEs and a discrete-state Markov jump process that, taken together, form an inexact yet computationally amenable representation of a well-mixed CRN. The key idea behind the method is to run, simultaneously, a numerical method for solving the system of ODEs alongside a SSA for simulating stochastic trajectories. Individuals in the system are represented by exactly one of the two regimes at any given time, but are permitted to switch back and forth between each modelling regime in response to the current concentration of their species. To accomplish this, we describe a `network extension' procedure by which one can convert a CRN into a larger network that is probabilistically equivalent to the original in a manner that we describe. The extended network is larger than the original in three specific ways. First, each species in the original corresponds to two species in the extended network, where one species is to be governed by the discrete regime and the other by the continuous. Second, to satisfy the combinatorial requirements that give rise to the probabilistic equivalence of each network, the extension requires that we add additional reactions that allow the continuous and discrete species to interact. The final ingredient in the extended network are first-order conversion reactions that allow discrete species to enter the continuous regime and vice versa, adaptively redistributing species concentrations between regimes to maximise computational efficiency and accuracy.

  From the extended network we construct an \textit{augmented reaction network} (ARN) that governs the same species as the extended network. The critical difference is that we represent the species marked as `continuous` (and the reactions between them) in the extended network by system of ODEs. This system of ODEs is derived by forming the CME that would govern the continuous species (were they discrete) from the set of reactions that act \textit{exclusively} on continuous species, deriving the moment equations for these species, and taking an appropriate moment closure. Under this representation, reactions between continuous species are governed exclusively by the continuum approximation, and reactions between discrete species are governed exclusively by the discrete simulation regime. To retain accuracy in bimolecular reactions, and to mitigate the impact of moment closure, reactions that have both a continuous and a discrete reactant are governed by the discrete simulation regime. Given that mass is converted back-and-forth between discrete and continuous representations depending on copy-number, we can reasonably view the ARN as a mechanism for representing `low copy-number reactions' under the discrete simulation regime, and `high copy-number reactions' under the continuum approximation. This new structure, the ARN, provides an intermediate description of a CRN that is both continuous and discrete. The MCM, then, is a method for simulating the trajectories of an ARN. We find that the MCM can indeed strike a balance between efficiency and accuracy.

  This remainder of this work is divided into three sections. In Section 2, we outline the construction of an ARN from a CRN alongside the mathematical prerequisites, the theoretical justification, and the specific algorithmic formulation of the MCM. In Section 3, we present numerical results that evaluate the accuracy and bias of our method for a series of test problems of increasing complexity. We conduct this evaluation by comparing the results from the MCM against results from an exact SSA. Finally, in Section 4 we give remarks on the relative advantages and limitations of our method versus traditional stochastic or numerical methods, and signpost future potential avenues of development and application for the method.

\section{Method} %!%--2 Method--%%%%%%%%%%%%%%%%%%%%%%%%%%%%%%%%%%%%%%%%%%%%%%%%%%%%%%%%%%%%%%%%%%%%%%%%%%%%%%%%%%%%%%%%
  In this section we describe the mass-conversion method (MCM) which couples a CRN described by a discrete-state Markov jump process with a system of ordinary differential equations representing the mean dynamics of the same CRN. We begin our discussion of the method with some preliminary information regarding stochastic simulation and continuum modelling before presenting the theoretical justification and implementation of our proposed coupling scheme.

  \subsection{Stochastic simulation and stoichiometry} %!%--2.1 Stochastic Simulation-----------------------------------
    We consider a CRN, $\sN$, with $K$ chemical species that interact via a set $R$ of reaction channels within a reaction vessel of unit volume. Denote by $X_k(t) \in \mathbb N$, for $k=1,\hdots,K$, the number of individuals of the $k^\text{th}$ species at continuous time $t$, and denote the overall state of the system by $\bm X(t) := (X_1(t),\hdots,X_K(t))$. We make the assumption that reaction $r\in R$ fires with an exponentially distributed waiting time with rate $\lambda_r$. The reaction rate coefficient $\lambda_r$ is typically taken to be constant over time; however, we note that the results in the remainder of this paper hold in the case that $\lambda_r$ is piecewise constant in time, with the caveat that there are only finitely many such discontinuities. Reactions in the network take the form
    \begin{equation}
      \sum_{k=1}^K \mu_{rk} X_k \xrightarrow{\lambda_r} \sum_{k=1}^K \eta_{rk} X_k,\quad\text{for } r \in R,
    \end{equation}
    where $\bm \mu_r = (\mu_{rk})_{k=1,\hdots,K}$ and $\bm \eta_r = (\eta_{rk})_{k=1,\hdots,K}$. We can thus, for each reaction, define the stoichiometric vector
    \begin{equation}
      \bm \nu_r := \bm \eta_r - \bm \mu_r
    \end{equation}
    which represents the change in state upon the firing of reaction $r$. These vectors are often collected into a single stoichiometric matrix, which we denote $\mathbf S$, where each column in $\mathbf S$ corresponds to a stoichiometric vector $\bm \nu_r$. To form this matrix, one must decide on an ordering of the reactions in $R$ - we note that this choice is arbitrary and bears no impact on the dynamics of the system.
    
    The most common method for drawing sample trajectories of $\bm X(t)$ is the aforementioned Gillespie direct method (GDM). Whilst the coupling technique for our hybrid method, which we will discuss later, is strictly independent of the choice of SSA, we will describe its implementation under the Gillespie direct method.
  
  \subsection{Continuum modelling} %!%--2.2 Continuum Modelling---------------------------------------------------------
    Given a CRN, $\sN$, we can derive the associated CME as follows. Define for each reaction a propensity function $\alpha_r(\bm X(t))$, defined such that  $\alpha_r(\bm X(t))\text dt$ is the probability that said reaction occurs within the infinitesimally small time interval $[t, t + \text dt)$. Under the law of mass-action, the propensity functions are given by
    \begin{equation}
      \alpha_r(\bm x) := \lambda_r \prod_{k=1}^K \frac{x_k!}{(x_k - \mu_{rk})!},
    \end{equation}
    where for brevity we have subsumed any combinatorial coefficients into the rate coefficient $\lambda_r$ \cite{van_kampen_stochastic_2007}. Standard techniques \cite{gillespie_rigorous_1992} reveal that the corresponding CME for this system is given by
    \begin{equation}
    \frac{\text d p(\bm x, t)}{\text d t} = \sum_{r \in R} \left[ \alpha_r(\bm x - \bm v_r) p(\bm x - \bm v_r, t) - \alpha_r(\bm x) p(\bm x,t) \right],\label{eq:CME}
    \end{equation}
    where $p(\bm x,t)$ is the probability that $\bm X(t) = \bm x$ at time $t$. Multiplying Equation~\eqref{eq:CME} by $x_k$ and summing over the state space $x_k$, yields the evolution equation for the mean concentration of each species. Denoting by $\langle f(\bm x) \rangle$ the expectation of $f(\bm x)$ with respect to $p(\bm x, t)$ for some function $f$, we have
    \begin{equation}
      \frac{\text d \langle x_i \rangle}{\text d t} = \sum_{r \in R} \nu_{ri} \langle \alpha_r(\bm x) \rangle. \label{eq:first-order-moments-eq}
    \end{equation}
    Defining the vector of propensity functions $\bm \alpha(\bm x) = (\alpha_r(\bm x))_{r \in R}$, this can be written in matrix form,
    \begin{equation}
      \frac{\text d \langle \bm x \rangle}{\text d t} = \mathbf S \langle\bm  \alpha(\bm x) \rangle,
    \end{equation}
    assuming that the enumeration of reactions in the vector $\bm \alpha$ corresponds to the column order of matrix $\mathbf S$. One can likewise, albeit through a somewhat laborious calculation, obtain higher-order moments of the system.
    %; namely
    % \begin{align*}
    %   \frac{\text d}{\text d t} \left\langle \prod_{i=1}^K x_i^{n_i} \right\rangle &= \sum_{r \in R} \left\langle \alpha_{r}(\bm x) \prod_{i=1}^K (x_i + \nu_{ri})^{n_i}\right\rangle \\
    %   &\quad - \sum_{r \in R} \left\langle \alpha_{r}(\bm x) \prod_{i=1}^K x_i^{n_i} \right\rangle,
    % \end{align*}
    %where $n_i \in \mathbb N_0$ for $i=1,...,K$. 
    These equations, however, do not in general admit closed-form solutions. Indeed, for CRNs with reactions of at least second-order, the system of moment equations itself is not closed; for example, for species which are reactants in a second-order reaction, the equation governing the evolution of the first moment of that species depends on the equations for the second moments, the equations for the second moments depend on the equations for the third moments, and so on.
    
    Making a moment-closure approximation requires the explicit adoption of some set of assumptions about the moments of a system. As such, these closures are necessarily \textit{ad hoc} and it is, in general, impossible to quantify a given closure's accuracy \textit{a priori}. Nevertheless, there are several closures that see wide application. The simplest and possibly most common closure is the so-called `mean-field' closure \cite[p.~82]{barrat_dynamical_2008}. Under the mean-field closure, all second-order central moments are assumed to be zero, yielding
    \begin{equation}
      \lr[]{x_i x_j} = \langle x_i \rangle \langle x_j \rangle,
    \end{equation}
    for all $i,j = 1,\hdots,K$. Another common closure is the Poisson closure \cite{nasell_extension_2003}, which assumes that variances are equal to their corresponding means and that all covariances are zero.
    \begin{equation}
      \lr[]{x_i^2} = \langle x_i \rangle + \langle x_i \rangle^2 \enspace\text{and}\enspace \langle x_i x_j \rangle = \langle x_i \rangle \langle x_j \rangle,
    \end{equation}
    for all $i,j=1,\hdots,K$ where $i\neq j$. Both the mean-field and Poisson closures close the system of moment equations at first-order. While there exist several higher-order closures \cite{schnoerr_comparison_2015}, they are generally unsuitable for use in hybrid methods, as there is currently no clear method for coupling higher-order moment equations to SSAs.
  
  \subsection{Reaction network extension} %!%--2.3 extension------------------------------------------------------
  \label{sec:extension}
  
    We begin our discussion of the MCM by noting that we will henceforth only consider reactions of at most second-order. These are reactions for which at most two individual reactant molecules are present. While a simultaneous interaction of three or more individuals is, in principle, possible, collision theory suggests that the probability of three or more distinct molecules interacting simultaneously is vanishingly small (see, e.g. \cite{laidler_reaction_1963}). Accordingly, a more realistic description of interactions of this type involves the formation of a highly reactive intermediary complex that subsequently reacts with the remaining reactants --- such a system is of at most second order~\cite{aris_modelling_1988}.
    
    The MCM partitions each chemical species $X_k$ into two `partition species', $C_k$ and $D_k$, each of which is governed by a different modelling regime, termed \textit{continuous} and \textit{discrete}, respectively. On these extension species we define a new reaction network that is both equivalent to the original network and computationally amenable. Further, this new `extended' reaction network contains additional `conversion' reactions that permit individuals to switch their partition at a rate proportional to the species-wise density. To do so, for each reaction in the network, we generate a new extended set of reactions for each possible combination of reactant regimes. In each reaction $r$, at most two species appear as reactants, which we label without loss of generality $X_i$ and $X_j$, where $i, j \in \{1,\hdots,K\}$, and where we may have that $i=j$. We require that this extended set of reactions obeys the following criteria:

    \begin{enumerate}[label=C\arabic*]
      \item To maximise efficiency, we wish to minimise unnecessary conversion back-and-forth between regimes. We thus determine that all molecules produced by reaction $r$ belonging to the $i^\text{th}$ species (resp. $j^\text{th}$) are allocated to the same regime as reactant $X_i$ (resp. $X_j$). \label{hypothesis-1}
      \item To maximise accuracy, we aim to retain much of the stochasticity in the system. In particular, for each reaction $r$, we allocate all product molecules from non-reactant species (i.e. species other than $X_i$ and $X_j$) to the discrete regime. \label{hypothesis-2}
      \item Applying \ref{hypothesis-2} without further restriction could yield a `trivial' reaction network wherein all continuous molecules are gradually converted to discrete molecules over time. As such, for reactions $r$ where all reactant molecules are in the continuous regime, we assign all the reaction's products to the continuous regime also. \label{hypothesis-3}
    \end{enumerate}
    
    We begin our exposition of the MCM with reactions of order zero; that is, reactions of the form
    \begin{equation}
      \emptyset \rightarrow P
    \end{equation}
    for some set of reaction products $P$. The choice of whether to place these reaction products into the discrete or continuous regime may be problem dependent; specifically, it may be the case that all products in $P$ belong to species that are known \textit{a priori} to be of high copy number, and as such might best be placed in the continuous regime. Nevertheless, in light of~\ref{hypothesis-2}, we place any such products into the discrete regime.
    
    First-order reactions are dealt with trivially when applying the criteria above. Specifically, reactions of the form
    \begin{equation}
      X_i \xrightarrow{\lambda_r} \sum_{k=1}^K \eta_{rk} X_{k}, \label{eq:order1}
    \end{equation}
    are extended into
    \begin{align}
      C_i &\xrightarrow{\lambda_r} \sum_{k=1}^K \eta_{rk} C_k, \label{eq:order1c}\\
      D_i &\xrightarrow{\lambda_r} \sum_{k=1}^K \eta_{rk} D_k, \label{eq:order1d}
    \end{align}
    Any second-order reaction $r\in R$ can be written uniquely in the form
    \begin{equation}
      X_i + X_j \xrightarrow{\lambda_r} \eta_{ri}X_i + \eta_{rj}X_j + \sum_{\substack{k=1,\hdots,K \\ k\neq i,j}} \eta_{rk} X_k, \label{eq:order2}
    \end{equation}
    for some $i, j \in \{1,\hdots,K\}$ with $i\leq j$. To extend such a reaction we consider the four possible combinations of reactant regimes and apply~\ref{hypothesis-1}~-~\ref{hypothesis-3}, yielding
    \begin{align}
      C_i + C_j &\xrightarrow{\lambda_r} \sum_{k=1}^K \eta_{rk} C_k, \label{eq:order2cc} \\
      D_i + C_j &\xrightarrow{\lambda_r} \eta_{rj} C_j + \sum_{\substack{k=1,\hdots,K\\k\neq j}} \eta_{rk} D_k, \label{eq:order2dc}\\
      C_i + D_j &\xrightarrow{\lambda_r} \eta_{ri} C_i + \sum_{\substack{k=1,\hdots,K\\k\neq i}} \eta_{rk} D_k, \label{eq:order2cd}\\
      D_i + D_j &\xrightarrow{\lambda_r} \sum_{k=1}^K \eta_{rk} D_k,\label{eq:order2dd}
    \end{align}
    Note that in the case of a homodimerisation, where $i=j$, reactions~\eqref{eq:order2dc}~and~\eqref{eq:order2cd} are identical. Nevertheless, both must be included in the resultant network --- this is explained in detail in Section~\ref{sec:equivalence}. Applying this extension procedure to each reaction in the original network yields a new extended reaction network with chemical species $C_k$ and $D_k$ for $k = 1,\hdots,K$.
    
    Remaining are the regime conversion reactions that facilitate the conversion of species at high- and low-copy numbers to the continuous and discrete regimes, respectively. To this end, we append to the extended network reactions of the form
    \begin{equation}
      C_k \xrightleftharpoons[\kappa_{b,k}]{\kappa_{f,k}} D_k,
    \end{equation}
    where $\kappa_{f,k}$ and $\kappa_{b,k}$ are non-constant rates of the form
    \begin{align}
    \begin{split}
      \kappa_{f,k} &\dfn \gamma_{f,k} \mathbf 1_{\{C_k + D_k < T_{f,k}\}}, \\
      \kappa_{b,k} &\dfn \gamma_{b,k} \mathbf 1_{\{C_k + D_k > T_{b,k}\}},
    \end{split}
    \label{eq:conversion-reacs}
    \end{align}
    for pre-determined regime-conversion rates $\gamma_{f,k}$ and $\gamma_{b,k}$, conversion thresholds $T_{f,k}$ and $T_{b,k}$, and where the subscript characters $f$ and $b$ indicate the `forward' and `backward' conversions respectively. The collection of the species $C_k$ and $D_k$ for $k=1,\hdots,K$ and the set of reactions obtained from the procedures detailed above form the extended version of the network $\sN$. In completing our description of this network, it is useful at this point to introduce notational conventions that reflect both its structure and its provenance. For a CRN $\sN$, we denote its extended version by $\widetilde \sN$. We denote the state vector of $\wsN$ by $\bm Y(t)$, taking without loss of generality $\bm Y(t) \dfn \bm{C}(t) \oplus \bm{D}(t)$, where $\bm{C}(t) = (C_1, \hdots, C_K)$, $\bm{D}(t) = (D_1, \hdots D_K)$, and the operator $\oplus$ denotes vector concatenation. Finally, we denote the collection of reactions in $\wsN$ by $\widetilde R$.
  
  \subsection{Network equivalence} %!%--2.4 Network Equivalence---------------------------------------------------------
  \label{sec:equivalence}
  
    We claim that the evolution of the quantity $X_k$ in the CRN $\sN$ is the same as the evolution of the quantity $C_k + D_k$ in the paritioned version $\widetilde\sN$, for all $i=1,\hdots,K$, provided that the species $C_k$ are treated as discrete and simulated using the stochastic simulation algorithm. Before embarking on the derivation of this equivalence, we must first specify what, precisely, we are aiming to demonstrate. Define $p(\bm x,t)$ to be the probability that $\{\bm X(t) = \bm x\}$ and $q(\bm x,t)$ to be the probability that $\{\bm{C}(t) + \bm{D}(t) = \bm x\}$. Our aim, therefore, is to demonstrate that for any choice of $\bm x \in \mathbb N^K$ and $t>0$ we have $q(\bm x,t) = p(\bm x, t)$, provided that the initial conditions for $C_k + D_k$ are the same as those for $X_k$.

    To this end, consider a CRN $\mathcal N$ with $K$ species and $|R| = |R_0| + |R_1| + |R_2|$ reactions, where $R_0$, $R_1$, and $R_2$ are the sets of zeroth-, first-, and second-order reactions in the network, respectively. Recalling that the CME for this network is given by Equation~\eqref{eq:CME}, we rewrite the CME for $\sN$ in the form
    \begin{align}
    \begin{split}
      \frac{\text d}{\text dt} p(\bm x,t) &= \sum_{d=0}^2 \sum_{r\in R_d} \alpha_r(\bm x - \bm v_r) p(\bm x - \bm v_r, t) \\
      &\quad- \sum_{d=0}^2 \sum_{r\in R_d} \alpha_r(\bm x) p(\bm x,t).
    \end{split} \label{eq:master-equation}
    \end{align}
    
    The extension procedure from Section~\ref{sec:extension} gives a CRN $\wsN$ with $2K$ species and a set of reactions $\widetilde R$, where $|\widetilde R| = |R_0| + 2|R_1| + 4|R_2| + 2K$. We associate each reaction in $\widetilde{\sN}$ (excluding the $2K$ regime conversion reactions) with the original reactions from which they were extended. Each zeroth-order reaction in $\sN$ is associated with a zeroth order reaction in $\sN$. Similarly, first- and second-order reactions in $\sN$ are associated with two first- and four second-order reactions in $\widetilde{\sN}$, respectively. To track these relationships, we must introduce some new notation. We denote by $\widetilde{\bm \nu}_{r,\ell}$, where $r \in R_d$, $\ell=1,\hdots,2^d$, and $d=0,1,2$, the stoichiometric vectors for the $2^d$ reactions in $\widetilde{\sN}$ associated with reaction $r$ in $\sN$. In particular, notice that our extension procedure guarantees that
    \begin{equation}
      (\widetilde{\bm \nu}_{r,\ell})_{1:K} + (\widetilde{\bm \nu}_{r,\ell})_{K+1:2K} = \bm\nu_r,
      \label{eq:stoich-equality}
    \end{equation}
    for all reactions $r \in R$, $\ell=1,\hdots,2^d$, $d=0,1,2$, and where $\bm v_{n:m} = (v_n,...,v_m)$ for $n\leq m$. Additionally, we define the extended set of propensity functions for each reaction $r \in R$ via the usual mass-action kinetics, denoted by $\widetilde{\alpha}_{r,\ell}$ for $\ell=1,\hdots,2^d$. Note that in both cases, there is an implied ordering on the stoichiometric vectors and propensity functions associated with each reaction that is induced by $\ell$ - any such enumeration is arbitrary and exists only for notational utility; the only restriction is that the enumerations of stoichiometric vectors and propensity functions match for any given $r$.
    
    The propensity functions for the forward and backward regime conversion reactions~\eqref{eq:conversion-reacs} are not strictly governed by mass-action kinetics by virtue of their rates' dependence on the concentration of non-reactant species. Specifically, we choose the propensity functions for the forward and backward reactions for each species $k=1,\hdots,K$ to take the forms
    \begin{align*}
      \widetilde{\alpha}_{f,k}(\bm y) &\stackrel{\text{def}}{=} \gamma_{f,k} d_k \mathbf 1_{\{c_k + d_k > T_{f,k}\}}, \\
      \widetilde{\alpha}_{b,k}(\bm y) &\stackrel{\text{def}}{=} \gamma_{b,k} c_k \mathbf 1_{\{c_k + d_k < T_{b,k}\}},
    \end{align*}
    respectively, with associated stoichiometric vectors given by
    \begin{align*}
      \widetilde{\bm\nu}_{f,k} &\dfn \bm e_{k} - \bm e_{k+K},\\
      \widetilde{\bm\nu}_{b,k} &\dfn \bm e_{k+K} - \bm e_{k},
    \end{align*}
    again respectively, and where $\bm e_k$ denotes the $k^\text{th}$ standard basis vector in $\mathbb R^{2K}$. The CME for the network $\widetilde{\sN}$ can thus be expressed as
    \begin{align*}
      \frac{\text d}{\text dt} \widetilde{p}(\bm y, t) &= \sum_{d=0}^2  \sum_{\ell=1}^{2^d} \sum_{r\in R_d} \widetilde{\alpha}_{r,\ell}(\bm y - \widetilde{\bm\nu}_{r,\ell})\widetilde{p}(\bm y - \widetilde{\bm\nu}_{r,\ell},t) \\
      &\quad- \sum_{d=0}^2 \sum_{\ell=1}^{2^d} \sum_{r \in R_d} \widetilde{\alpha}_{r,\ell}(\bm y)\widetilde{p}(\bm y,t) \\
      &\quad+ \sum_{i=1}^K \widetilde\alpha_{f,i}(\bm y - \widetilde{\bm\nu}_{f,i}) \widetilde p(\bm y - \widetilde{\bm\nu}_{f,i},t) + \widetilde\alpha_{b,i}(\bm y - \widetilde{\bm\nu}_{b,i}) \widetilde p(\bm y - \widetilde{\bm\nu}_{b,i},t) \\
      &\quad- \sum_{i=1}^K \widetilde\alpha_{f,i}(\bm y) \widetilde p(\bm y,t) + \widetilde\alpha_{b,i}(\bm y) \widetilde p(\bm y,t),
    \end{align*}
    where $\widetilde{p}(\bm y, t)$ denotes the probability that $\{\bm Y(t) = \bm y\}$ at time $t$, where $\bm y = \bm{c} \oplus \bm{d}$. Recalling the definition of $q(\bm x, t)$, we can additionally write the master equation governing $q(\bm{c} + \bm{d}, t)$,
    \begin{align}
    \begin{split}
      \frac{\text d}{\text d t} q(\bm{c} + \bm{d}, t) &= \sum_{d=0}^2 \sum_{r \in R_d} q(\bm{c} + \bm{d} - \bm\nu_{r}, t) \sum_{\ell=1}^{2^d} \widetilde{\alpha}_{r,\ell}(\bm{c} \oplus \bm{d} - \widetilde{\bm\nu}_{r,\ell}^d)  \\
      &\quad- \sum_{d=0}^2 \sum_{r \in R_d} q(\bm{c} + \bm{d}, t) \sum_{\ell=1}^{2^d} \widetilde{\alpha}_{r,\ell}(\bm{c} \oplus \bm{d}),
    \end{split} \label{eq:summed-master-equation}
    \end{align}
    noticing that the regime conversion reactions contribute nothing to the evolution of $q$, since each conserves the quantity $\bm{c}(t) + \bm{d}(t)$. Comparing Equations~\eqref{eq:master-equation} and~\eqref{eq:summed-master-equation}, the critical step in our proof of equivalence is demonstrating that
    \begin{equation}
      \sum_{\ell=1}^{2^d} \widetilde{\alpha}_{r,\ell}(\bm{c} \oplus \bm{d}) = \alpha_r(\bm{x}), \label{eq:prove-this}
    \end{equation}
    for all $r \in R$, and for any $\bm{c}, \bm{d} \in \mathbb N^K$ where $\bm{c} + \bm{d} = \bm x$. To prove this, we will consider how the sum~\eqref{eq:prove-this} behaves for each reaction order. To begin, fix $\bm{c}, \bm{d} \in \mathbb N^K$ and $\bm x = \bm{c} + \bm{d}$. Consider the case $z=0$, where $z$ denotes the reaction order we are considering. For any zeroth order reaction under the law of mass-action, we trivially have that $\widetilde{\alpha}_{r,1}(\bm{c} \oplus \bm{d}) = \lambda_r = \alpha_{r}(\bm{c} + \bm{d})$ for all $r\in R_0$. Since each reaction $r \in R_0$ corresponds with exactly one reaction in $\widetilde R$, Equation~\eqref{eq:prove-this} holds for $d=0$.
    
    We now consider the case $z=1$ and consider a reaction $r\in R_1$ of the form~\eqref{eq:order1}, taking without loss of generality $\ell=1$ to denote the reaction~\eqref{eq:order1c} and $\ell=2$ to denote the reaction~\eqref{eq:order1d}. Notice that we have $\widetilde\alpha_{r,1}(\bm{c} \oplus \bm{d}) = \alpha_{r}(\bm{c})$ and $\widetilde\alpha_{r,2}(\bm{c} \oplus \bm{d}) = \alpha_{r}(\bm{d})$. Further, under mass-action, the functions $\alpha_r$ are linear for any first-order reaction $r$. Therefore, we have
    \begin{align*} 
    \widetilde\alpha_{r,1}(\bm{c} \oplus \bm{d}) + \widetilde\alpha_{r,2}(\bm{c} \oplus \bm{d}) &= \alpha_{r}(\bm{c}) + \alpha_{r}(\bm{d}) \\
    &= \lambda_r c_k + \lambda_r d_k = \lambda_r (c_k + d_k) \\
    &= \alpha_r(\bm{c} + \bm{d}) = \alpha_r(\bm x),
    \end{align*}
    for all $r \in R_1$ and Equation~\eqref{eq:prove-this} holds for first-order reactions. 
    
    Next, consider $z=2$ and consider a second-order reaction $r$ of the form~\eqref{eq:order2}. Similarly to the first-order case, we enumerate without loss of generality the propensity functions $\widetilde{\alpha}_{r,\ell}$ by setting $\ell=1,\hdots,4$ to correspond with reactions~\eqref{eq:order2cc} through \eqref{eq:order2dd}, respectively. Note that there are two distinct classes of second-order reaction; namely, homodimerisations, where both reactants are of the same species, and heterodimerisations, where both reactants are of different species. Each class yields propensity functions of a different functional form and must, therefore, be considered separately. For a homodimerisation $r$ of reactant species $X_k$, we have that
    \begin{align*}
      \widetilde{\alpha}_{r,1} (\bm{c} \oplus \bm{d}) &= \lambda_r(c_k^2 - c_k), \\
      \widetilde{\alpha}_{r,2} (\bm{c} \oplus \bm{d}) &= \lambda_r d_k c_k, \\
      \widetilde{\alpha}_{r,3} (\bm{c} \oplus \bm{d}) &= \lambda_r c_k d_k, \\
      \widetilde{\alpha}_{r,4} (\bm{c} \oplus \bm{d}) &= \lambda_r(d_k^2 - d_k), \\
      \intertext{under mass-action kinetics. Summing these four equations yields}
      \sum_{\ell=1}^4 \widetilde{\alpha}_{r,\ell} (\bm{c} \oplus \bm{d}) &= \lambda_r (c_k + d_k)(c_k + d_k - 1)= \alpha_r (\bm{c}+\bm{d}) = \alpha_r(\bm x),
    \end{align*}
    and therefore Equation~\eqref{eq:prove-this} holds for homodimerisations. Likewise, for a heterodimerisation $r$ and reactant species of reactant species $X_i$ and $X_j$, we have
    \begin{align*}
      \widetilde{\alpha}_{r,1} (\bm{c} \oplus \bm{d}) &= \lambda_r c_i c_j, \\
      \widetilde{\alpha}_{r,2} (\bm{c} \oplus \bm{d}) &= \lambda_r d_i c_j, \\
      \widetilde{\alpha}_{r,3} (\bm{c} \oplus \bm{d}) &= \lambda_r c_i d_j, \\
      \widetilde{\alpha}_{r,4} (\bm{c} \oplus \bm{d}) &= \lambda_r d_i d_j, \\
      \intertext{under mass-action kinetics. Summing these four equations yields}
      \sum_{\ell=1}^4 \widetilde{\alpha}_{r,\ell} (\bm{c} \oplus \bm{d}) &= \lambda_r (c_i + d_i)(c_j + d_j) = \alpha_r (\bm{c} + \bm{d}) = \alpha_r(\bm x),
    \end{align*}
    and therefore Equation~\eqref{eq:prove-this} holds for heterodimerisations. The final step of the proof is to observe that the innermost summand in the master equation~\eqref{eq:summed-master-equation} can be written
    \begin{align}
    \begin{split}
      \sum_{\ell=1}^{2^d}\widetilde{\alpha}_{r,\ell}(\bm{c} \oplus \bm{d} - \widetilde{\bm\nu}_{r,\ell}) &= \sum_{\ell=1}^{2^d}\widetilde{\alpha}_{r,\ell} \left ((\bm{c} - (\widetilde{\bm\nu}_{r,\ell})_{1:K}) \oplus (\bm{d} - (\widetilde{\bm\nu}_{r,\ell})_{K+1:2K}) \right ) \\
      &= \alpha_{r} \left((\bm{c} - (\widetilde{\bm\nu}_{r,\ell})_{1:K}) + (\bm{d} - (\widetilde{\bm\nu}_{r,\ell})_{K+1:2K}) \right) \\
      &= \alpha_{r} \left( \bm{c} + \bm{d} - \bm \nu_r \right) = \alpha_r(\bm x - \bm \nu_r),
    \end{split}\label{eq:final-equality}
    \end{align}
    where the second step follows from equivalence~\eqref{eq:prove-this} and the third follows from relationship~\eqref{eq:stoich-equality}. Taken together, Equations~\eqref{eq:prove-this}~and~\eqref{eq:final-equality} allow us to rewrite~\eqref{eq:summed-master-equation} as
    \begin{align*}
      \frac{\text d}{\text dt} q(\bm{c} + \bm{d}, t) = \frac{\text d}{\text dt} q(\bm x, t) &= \sum_{d=0}^2 \sum_{r \in R_d} \alpha_r(\bm x - \bm\nu_r,t) q(\bm x - \bm\nu_r, t) \\
      &\quad -\sum_{d=0}^2 \sum_{r \in R_d} \alpha_r(\bm x) q(\bm x, t), 
    \end{align*}
    which, upon inspection, is identical to the evolution equation that governs $p$; namely, Equation~\eqref{eq:master-equation}.
  
  \subsection{The augmented reaction network} %!%--2.5 The ARN----------------------------------------------------------
    In this subsection, we use the extended network $\widetilde{\sN}$ to construct an \textit{augmented reaction network} (ARN), which we denote $\sM$, that consists of both a chemical reaction network (simulated stochastically) and a set of ODEs (simulated deterministically) that, taken together, provide an approximation of the original network $\sN$ and that can be simulated at lower computational expense. Indeed, simply simulating the network $\widetilde{\sN}$ using an SSA would be at least as computationally expensive as simply simulating $\sN$. Specifically, the ARN contains all $2K$ species of $\widetilde\sN$ --- the key difference is that in forming the ARN we separate out all reactions that contain only continuous species. These `continuous-only' reactions are not simulated using the discrete method; rather, we derive from the continuous-only reactions a system of approximate time-evolution equations that govern (in part) the means of the continuous species $C_k$. It is this system of equations that we simulate using the continuous method. Note that not \textit{all} reactions in which the $C_k$ participate are continuous-only; indeed, many of the first- and second-order reactions in $\widetilde\sN$ contain both continuous and discrete species. These reactions that involve both continuous and discrete species are of `mixed-type', and are simulated using the discrete method. In this manner, the discrete species are governed exclusively by the discrete method; on the other hand, the continuous species are governed by the continuous method for all high copy-number reactions (the continuous-only reactions) and by the discrete method for low copy-number reactions (the mixed-type reactions).
    
    We now detail the construction of the ARN. Beginning with a CRN, $\sN$, we apply the extension procedure set out in Section~\ref{sec:extension} to produce the extended network $\widetilde{\sN}$. As before, we denote by $\bm{C}(t)$ and $\bm{D}(t)$ the number of individuals in the continuous and discrete regimes at time $t$, respectively, which we combine into a single state vector $\bm Y(t) = \bm{C}(t) \oplus \bm{D}(t)$. The complete set of reactions in the extended network numbers $|R_0| + 2|R_1| + 4|R_2| + 2K$, of which a total of $|R_1| + |R_2|$ are continuously-only --- one for each first-order reaction and one for each second-order reaction in the original network. We denote the sets of continuous-only first- and second-order reactions by $R^c_1$ and $R^c_2$, respectively. From $R^c_1 \cup R^c_2$ we derive a master equation governing the evolution of $\mathbb P(\bm{C}(t)= \bm c(t))$ under this set of reactions. Finally, we derive mean time-evolution equations and close the system at first-order (via the mean-field or Poisson moment closures, for example). This procedure yields a system of ODEs that will ultimately be simulated by the continuous method. The remaining $|R_0| + |R_1| + 3|R_2| + 2K$ reactions are those aforementioned mixed-type and discrete-only reactions, which will be simulated by the discrete method.
    
    Following this procedure, we find that the mean of the $k^\text{th}$ continuous species under the action of the reactions in the set $R^c_1 \cup R^c_2$ obeys the following evolution equation,
    \begin{equation}
      \frac{\text d}{\text dt} \lr{C_i} = \sum_{r\in R^c_1 \cup R^c_2} \nu_{ri} \lr{\alpha_r(\bm{c}(t))}. \label{eq:cts-means}
    \end{equation}
    Given that this description contains only first and second-order reactions, it is straightforward to derive mean time-evolution equations for each of the $C_i$ under the mean-field and Poisson closures. Define for a reaction $r$ the function $\pi_r(n)$ that returns the $n^\text{th}$ reactant species of said reaction, where $n=1,\hdots,d$. For example, for a reaction $r$ of the form~\eqref{eq:order2cc}, the function takes the values $\pi_r(1) = C_i$ and $\pi_r(2) = C_j$. Denote by $R^c_H$ and $R^c_O$ the sets of hetero and homodimerisations, respectively, such that $R^c_H \cup R^c_O = R^c_2$. Note that the definition of a homodimerisation guarantees that for any such reaction $r$, $\pi_r(1) = \pi_r(2)$. We can now write the mean time-evolution equations for each of the $C_k$. Under the mean-field closure, Equation~\eqref{eq:cts-means} becomes
    \begin{equation}
      \frac{\text d}{\text dt} \lr{C_k} = \sum_{r \in R^c_1} \lambda_r \nu_{rk} \lr{\pi_r(1)} + \sum_{r \in R^c_H} \lambda_r \nu_{rk} \lr{\pi_r(1)} \lr{\pi_r(2)} + \sum_{r \in R^c_O} \lambda_r \nu_{rk} \lr{\pi_r(1)}^2. \label{eq:arn-mean}
    \end{equation}
    Similarly, under the Poisson closure, Equation~\eqref{eq:cts-means} becomes
    \begin{equation}
      \frac{\text d}{\text dt} \lr{C_k} = \sum_{r \in R^c_1} \lambda_r \nu_{rk} \lr{\pi_r(1)} + \sum_{r \in R^c_H} \lambda_r \nu_{rk} \lr{\pi_r(1)} \lr{\pi_r(2)} + \sum_{r \in R^c_O} \lambda_r \nu_{rk}\left[ \lr{\pi_r(1)} + \lr{\pi_r(1)}^2 \right]. \label{eq:arn-poisson}
    \end{equation}

    To complete our description of the ARN, we also must specify the stoichiometry matrix, denoted $\mathbf M$, that represents the set of reactions that will be simulated using the discrete method. This matrix may be written in block form,
    \begin{equation}
      \mathbf M = \begin{bmatrix} \mathbf M_R & \mathbf M_K \end{bmatrix},
    \end{equation}
    where $\mathbf M_R$ is the stoichiometric matrix obtained all remaining $|R_0| + |R_1| + 3|R_2|$ discrete reactions in $\widetilde{\sN}$, and $\mathbf M_K$ is the stoichiometric matrix representing the regime conversion reactions. Notice that without loss of generality we can write
    \begin{align*}
      \mathbf M_K =  \begin{bmatrix}
      \mathbf I_K & -\mathbf I_K \\ -\mathbf I_K & \mathbf I_K
      \end{bmatrix},
    \end{align*}
    where $\mathbf I_K$ is the $K \times K$-dimensional identity matrix.
    
    The ARN corresponding to the CRN $\sN$ is thus defined to be the tuple of the set of $2K$ species $C_i, D_i$ ($i=1,\hdots,K$), the stoichiometry matrix $\mathbf M$ and associated propensity functions $\widetilde\alpha_{r,\ell}$ ($r \in R_d, k = 1,\hdots,2^d, d=0,1,2$), and the system of ODEs given by either~\eqref{eq:arn-mean} or~\eqref{eq:arn-poisson}, depending on the chosen closure. We call these the mean-field ARN (M-ARN) and the Poisson ARN (P-ARN) associated with the CRN $\mathcal N$, respectively.

  \subsection{The Mass-Conversion Method} %!%--2.6 The MCM--------------------------------------------------------------
    We now describe in detail our proposed algorithm for the efficient simulation of an ARN $\sM$: the mass-conversion method. The method itself resembles that of other hybrid methods based on the Gillespie direct method, and its implementation is straightforward --- the mathematical machinery that gives the method its computational efficiency is implicit in the structure of the ARN.
    
    The only strictly numerical parameter in the method is $\Delta t$, the ODE update step size, which should be chosen according to the numerical method used for solving the system of ODEs. In the present description of the method, we take this step size $\Delta t$ to be fixed; however, we note that all instances of fixed $\Delta t$ may be replaced with a suitable value to accommodate, for example, adaptive time-stepping methods.
    
    The method is initialised by specifying the initial conditions $\bm Y(0) = \bm{C}(0) \oplus \bm{D}(0)$, the first ODE update time, $t_d = \Delta t$, and the initial and final simulation times $t_0$ and $t_f$, respectively. We next calculate the value of each propensity function at the initial time $t = t_0$ and calculate their sum $\alpha_0(t)$. As in the Gillespie direct method, the sum $\alpha_0(t)$ is used to determine the time until the next discrete-regime reaction $\tau$ using the formula
    \begin{equation*}
      \tau = \frac{1}{\alpha_0}\ln\left( \frac{1}{u} \right),
    \end{equation*}
    where $u \sim U(0,1)$ is a uniformly distributed random number.
      
    If, at time $t$, the time of the next reaction is before that of the next ODE update (i.e. $t + \tau < t_d$) then a regular stochastic event is executed. Notice, however, that since the state $\bm{C}$ is partially governed by the system of ODEs, the mass of any given species $C_k$ is not nececssarily integer-valued. It is possible then that the firing of an event in the usual manner may result in $C_k < 0$ for some $k=1,\hdots,K$. To avoid this unphysical occurence we perform a rejection sampling step when a reaction attempts to destroy or convert a continuous mass molecule of species $k$ when $C_k \in (0,1)$. Specifically, we sample $u \sim U(0,1)$ -- if $u < C_k$, we execute the reaction and set $C_k = 0$; otherwise, the reaction does not occur.
    
    If $t + \tau > t_d$ we set $t = t + \tau$. Then, we enumerate without loss of generality all reactions by the order in which they appear in the stoichiometry matrix $\mathbf M$ of $\sM$, denoting by $\widetilde{\alpha}_p(t)$ the value of the propensity function at time $t$ associated with the $p^\text{th}$ reaction under said enumeration. The reaction to be executed is then sampled by selecting $r \sim U(0,1)$ uniformly at random and finding $j$ such that 
    \begin{equation*}
      \sum_{p=1}^j \widetilde{\alpha}_p(t) < r \alpha_0 < \sum_{p=1}^{j+1} \widetilde{\alpha}_p(t).
    \end{equation*}
    In the case that the next reaction would occur after that of the next ODE update (i.e. $t + \tau > t_d$), an ODE update is performed to calculate the concentrations of the continuous species $\bm{C}$. This may be achieved using any suitable numerical method. After this, the time is set to be equal to the current ODE update time $t=t_d$, the time of the next ODE update is set $t_d = t_d + \Delta t$, and the process of sampling a new stochastic event is begun anew at time $t$. This procedure continues until the final time $t_f$ is reached, and forms the entirety of the MCM. An algorithmic description of the MCM is given in Algorithm~\ref{alg:mcm} 

\section{Results} %!%--3 Results--%%%%%%%%%%%%%%%%%%%%%%%%%%%%%%%%%%%%%%%%%%%%%%%%%%%%%%%%%%%%%%%%%%%%%%%%%%%%%%%%%%%%%%
\label{sec:results}
  In this section we demonstrate the accuracy of the MCM for three example problems of increasing complexity. We choose to use the classical fourth-order Runge Kutta method~(see, e.g.~\cite[p.~352]{suli_introduction_2003}) for solving the systems of ODEs, and the GDM for simulating stochastic trajectories. We make special note that the validity of our coupling is independent of the chosen numerical method for simulating the system of ODEs. Nevertheless, the accuracy of the method as a whole will naturally depend to a large extent on the accuracy of the underlying numerical techniques; a phenomenon that we explore in Test Case~\ref{sec:test-case-2}. To measure the error in a simulation run, we define the \textit{relative error} between the SSA and the MCM by
  \begin{equation}
      \varepsilon_{k,\text{MCM}}(t) \dfn \frac{f_{k,\text{MCM}}(t) - f_{k,\text{SSA}}(t)}{f_{k,\text{SSA}}(t)},
  \end{equation}
  where $f_{k,\text{SSA}}$ is the computed density of the $k^\text{th}$ species at time $t$ as approximated by the SSA (resp.\ by the MCM). Likewise, we define the relative error between the system of ODEs and the SSA by
  \begin{equation}
      \varepsilon_{k,\text{ODE}} \dfn \frac{f_{k,\text{ODE}}(t) - f_{k,\text{SSA}}(t)}{f_{k,\text{SSA}}(t)},
  \end{equation}
  where $f_{k,\text{ODE}}$ is the computed density of the $k^\text{th}$ species at time $t$ according to the system of ODEs as simulated by the numerical method.

  \subsection{Test Case 1 --- Alternating exponential growth} %!%--3.1 altexp-------------------------------------------
\begin{figure}[htbp]
      \centering
      \subfloat[Density plot]{\includegraphics[width=80mm]{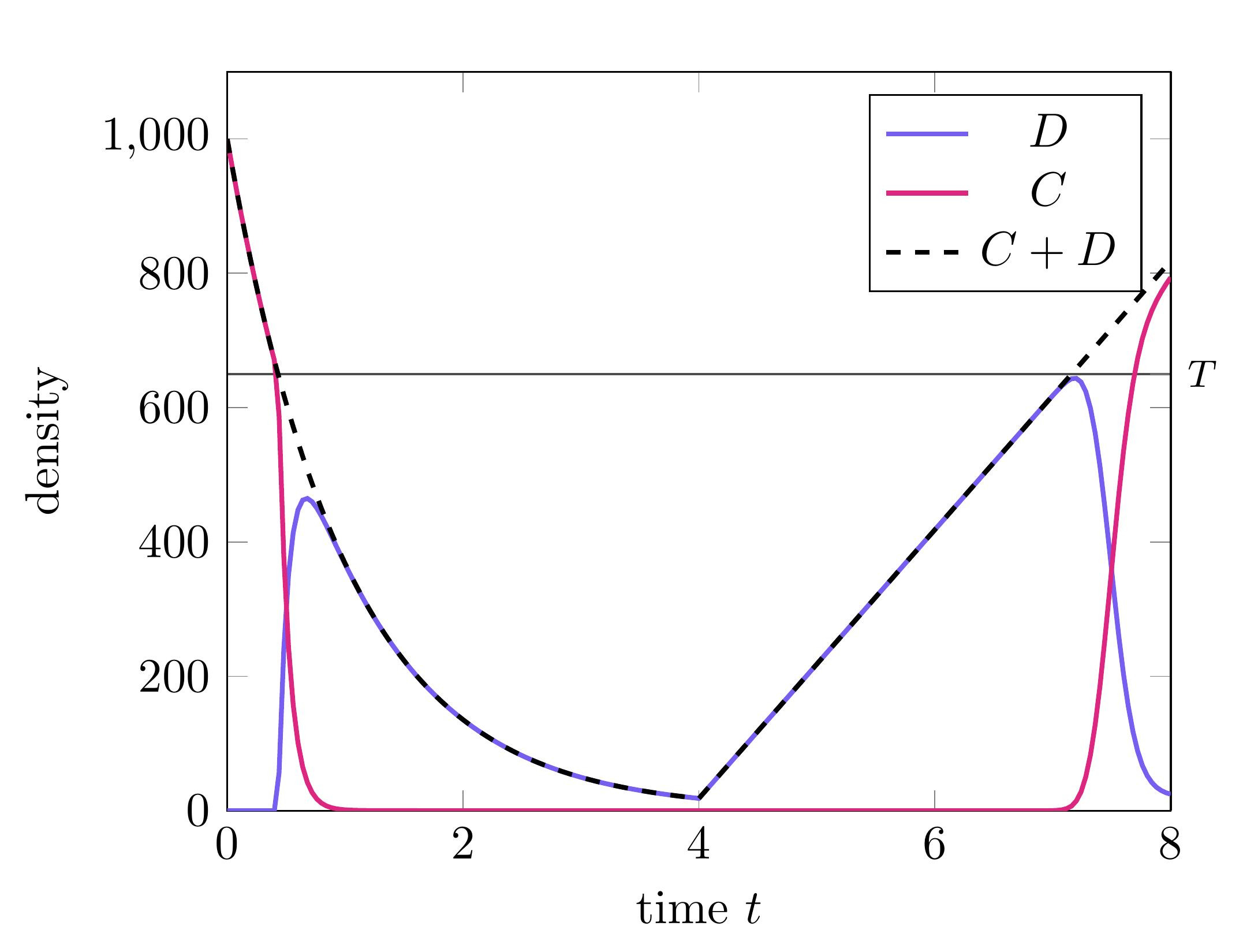}}%
      \subfloat[Relative error plot]{\includegraphics[width=80mm]{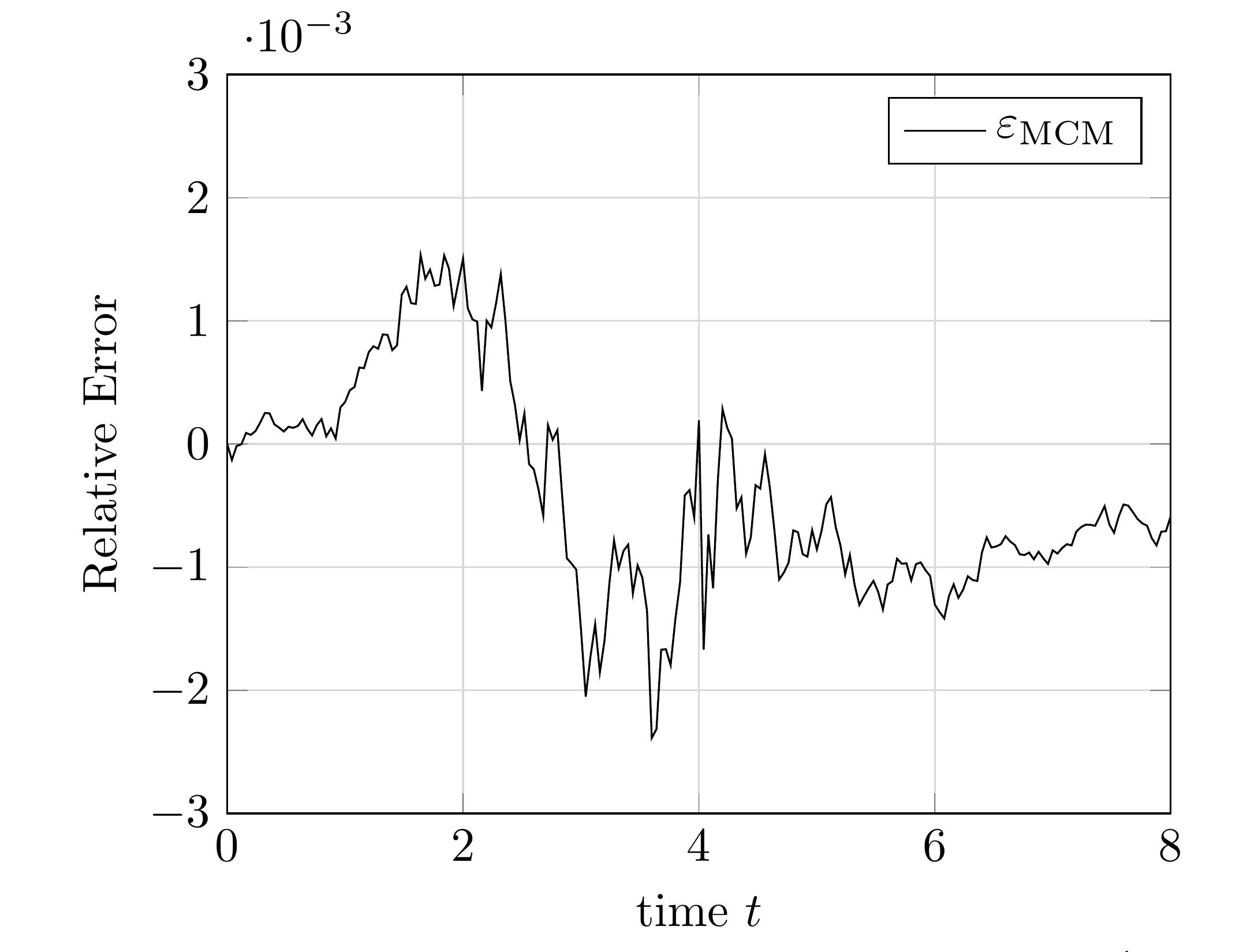}}
      \caption{Results of Test Case 1 (Section~\ref{sec:test-case-1}). (A) Plot of the density of $X$ as simulated by the MCM with parameters as specified in Table~\ref{table:test-1-params} with conversion threshold $T=650$. (B) Relative error in $X$ between the MCM and the SSA. Simulation results averaged over $10^5$ repeats.}\label{fig:test-1-1}
    \end{figure}

  \label{sec:test-case-1}
    Our first test case aims to demonstrate the accuracy of the method in the case of network with a single species, where continuous mass is degraded by a first-order degradation reaction to induce a continuous-to-discrete regime conversion, and discrete mass is produced by a zeroth-order production reaction to induce a discrete-to-continuous regime conversion. We thus consider the following simple reaction network $\sN$ consisting of a single species $X$ and two reactions,
    \begin{equation}
        \emptyset \xrightleftharpoons[\lambda_2]{\lambda_1} X,
    \end{equation}
    where the rates are of the form 
    \begin{equation} 
        \lambda_i(t) = \begin{cases}
            k_{i} & t \in I_i, \\
            0 & \text{otherwise,}
        \end{cases}
    \end{equation}
    where $k_i > 0$ and $I_i$ is some finite, non-empty union of time intervals. In our specific example, we choose these intervals such that the degradation reaction is `on' precisely when the production reaction is `off', and vice-versa. This network has stoichimetry matrix
    \begin{equation}
        \mathbf S = \begin{bmatrix}
        1 & -1
        \end{bmatrix},
    \end{equation}
    and propensity functions
    \begin{equation*}
        \begin{array}{cc}
            \alpha_1 = \lambda_1, & \alpha_2 = \lambda_2 x.
        \end{array}
    \end{equation*}
    From this, we form the corresponding M-ARN $\sM$ with two species $C$ and $D$. This network has stoichimetry matrix
    \begin{equation*}
        \mathbf M = \begin{bmatrix}
        \begin{array}{rrr}
            1 & 0 \\
            0 & -1
        \end{array} & \rvline{} & \mathbf{M_1}
        \end{bmatrix}, \quad \text{recalling } \mathbf M_1 = \begin{bmatrix}
          1 & -1 \\
          -1 & 1
        \end{bmatrix},
    \end{equation*}
    and propensity functions
    \begin{equation*}
        \begin{array}{lll}
        \widetilde{\alpha}_{1,1} = \lambda_1, & \widetilde{\alpha}_{2,1} = \lambda_2 d, \\
    \end{array}
    \end{equation*}
    corrsponding to the zeroth-order production and the first-order degradation of discrete mass, respectively. The first-order degradation of continuous mass is modelled via the ODE
    \begin{equation}
        \frac{\text d}{\text d t} \lr[]{C} = -\lambda_2 \lr[]{C}.
    \end{equation}
    We present the results of this test case in Figure~\ref{fig:test-1-1} using the parameter values given in Table~\ref{table:test-1-params}. This proof-of-concept example demonstrates the key behaviour of the MCM --- the conversion between discrete- and continuum-governed mass. As expected, when overall density falls below the threshold value we observe the conversion of continuum to discrete mass, and vice versa when density again becomes sufficiently high. We observe no evidence of bias in the MCM, with the fluctuations away from zero in Figure~\ref{fig:test-1-1} not persisting between simulation runs.
    
  \subsection{Test Case 2 -- Alternating logistic growth} %!%--3.2 altlog-----------------------------------------------
  \label{sec:test-case-2}
    Our second Test Case aims to demonstrate the accuracy of the method in the case of a network with a single species, this time where continuous mass is degraded by a \textit{second}-order degradation reaction to induce a continuous-to-discrete regime conversion, and discrete mass is produced by a \textit{first}-order production reaction to induce a discrete-to-continuous regime conversion. As in Test Case 1, we take \(\sN\) consisting of a single species \(X\), this time with reactions
    \begin{equation}
        X \xrightleftharpoons[\lambda_2]{\lambda_1} X+X,
    \end{equation}
    where the rate $\lambda_1$ is constant over time and $\lambda_2$ is governed by
    \begin{equation} 
        \lambda_2(t) = \begin{cases}
            k_{2} & t \in I, \\
            0 & \text{otherwise,}
        \end{cases}
    \end{equation}
    where $\lambda_2 > 0$ and $I$ is some finite, non-empty union of time intervals. Again, we select these intervals such that the production reaction is `on' precisely when the degradation reaction is `off', and vice-versa. This network has stoichimetry matrix
    \begin{equation}
        \mathbf S = \begin{bmatrix}
        1 & -1
        \end{bmatrix},
    \end{equation}
    this time with propensity functions
    \begin{equation*}
        \begin{array}{cc}
            \alpha_1 = \lambda_1 x, & \alpha_2 = \lambda_2 x(x-1).
        \end{array}
    \end{equation*}
    Following extension, we obtain an ARN $\sM$ with two species $C$ and $D$. This network has stoichimetry matrix
    \begin{equation*}
        \mathbf M = \begin{bmatrix}
        \begin{array}{rrrr}
            1 & 0 & -1 & 0\\
            0 & -1 & 0 & -1
        \end{array} & \rvline{} & \mathbf{M_1}
        \end{bmatrix},
    \end{equation*}
    and propensity functions
    \begin{equation*}
        \begin{array}{lll}
        \widetilde{\alpha}_{1,1} = \lambda_1 d, & \widetilde{\alpha}_{2,1} = \lambda_2 d(d-1), & \walpha_{2,2}=\walpha_{2,3} = \lambda_2 dc, \\
    \end{array}
    \end{equation*}
    repesenting the production of a discrete molecule from a discrete molecule, the degradation of a discrete molecule by a discrete molecule, the degradation of a continuous molecule by a discrete molecule, and the degradation of a discrete molecule by a continuous molecule, respectively. We form the equation governing the second-order degradation of continuous mass by continuous mass and the production of continuous mass from continuous mass using the Poisson closure; this equation is given by the ODE
    \begin{equation}
        \frac{\text d}{\text dt} \lr[]{C} = \lambda_1\lr[]{C} - \lambda_2\lr[]{C}^2.
    \end{equation}
    
\begin{figure}[htbp]
    \centering
      \subfloat[Density plot]{%
        \includegraphics[width=80mm]{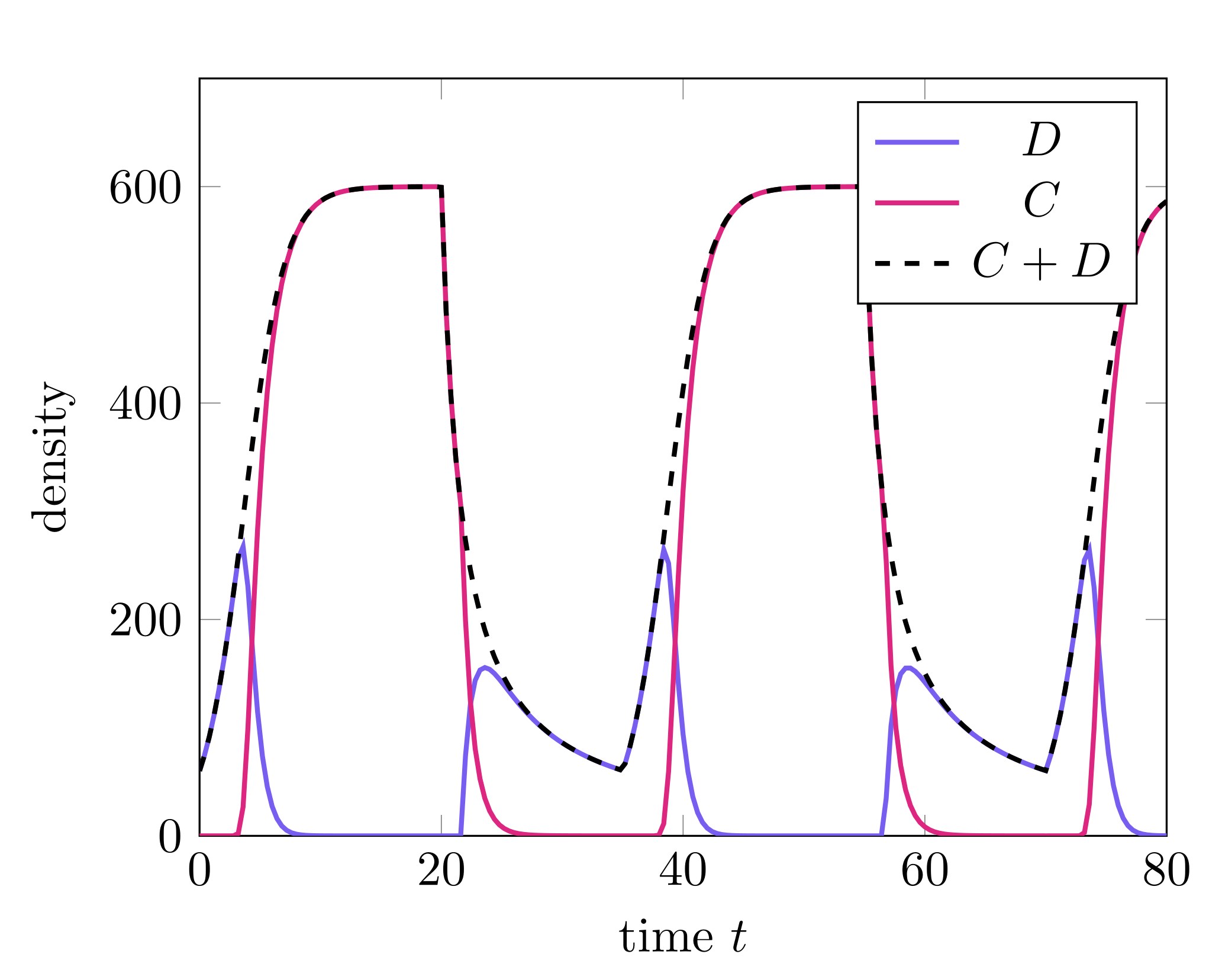}
      }
      \subfloat[Relative error plot]{
        \includegraphics[width=80mm]{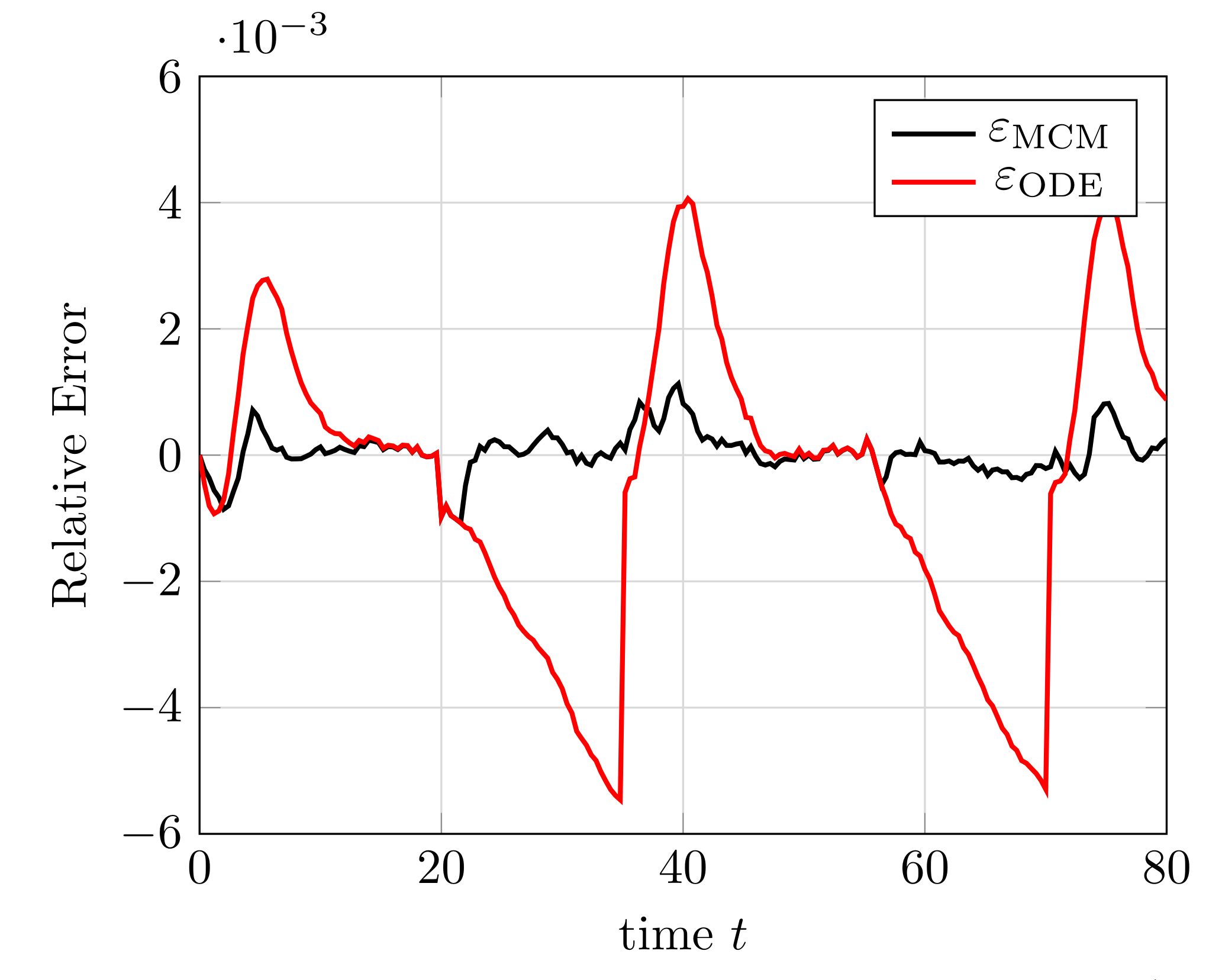}
      }
    \caption{Results of Test Case 2 (Section~\ref{sec:test-case-2}). (A) Plot of the density of $X$ as simulated by the MCM with parameters as specified in Table~\ref{table:test-1-params}. $C$ is shown in pink, $D$ is shown in blue, $C+D$ is shown dashed. (B) Relative error in $X$ between the MCM and the SSA. Simulation results averaged over $10^5$ repeats.}\label{fig:test-2-1}
    \end{figure}

    We present the results of this test case in Figure~\ref{fig:test-2-1} using the parameter values given in Table~\ref{table:test-1-params}. The results of this test case demonstrate a particular limitation of the MCM; namely, that the error in the MCM is, in some sense, `tethered' to the error in the solution to the system of ODEs in the associated ARN. We see this most clearly at the parameter transition point $t=20$, when the second-order reaction degradation activates.

  \subsection{Test Case 3 --- Chemical signalling} %!%--3.3 chmsig------------------------------------------------------
  \label{sec:test-case-3}
    For our third and final test case, consider a CRN, $\sN$, consisting of three chemical species $X_1, X_3,$ and $X_2$, which we refer to as the \textit{signal}, \textit{intermediate}, and \textit{product} species respectively, within a reactor vessel of unit volume. The product $X_2$ is produced via the intermediate $X_3$ and is degraded via a first-order sink reaction. The intermediate is produced via a zeroth-order source reaction.
    \begin{equation} 
        \label{eq:chemsig_1}
        \emptyset \xrightarrow{\lambda_1} X_3 \xrightarrow{\lambda_2} X_2 \xrightarrow{\lambda_3} \emptyset.
    \end{equation}
    The signal species $X_1$ is coupled indirectly with $X_2$ via the following reaction,
    \begin{equation} 
        \label{eq:chemsig_2}
        X_1 + X_3 \xrightarrow{\lambda_4} X_1, 
    \end{equation}
    in which the signal degrades the intermediate $X_3$. Finally, the signal species itself is produced and degraded according to the same reaction system we used in Test Case 1,
    \begin{equation} \label{eq:chemsig_3}
        \emptyset \xrightleftharpoons[\lambda_6]{\lambda_5} X_1.
    \end{equation}
    This CRN, $\sN$, has stoichimetry matrix
    \begin{equation*}
        \sS = \begin{bmatrix*}[r]
            0 & 0 & 0 & 0 & 1 & -1 \\
            1 & -1 & 0 & -1 & 0 & 0 \\
            0 & 1 & -1 & 0 & 0 & 0 
        \end{bmatrix*},
    \end{equation*}
    with propensity functions given by
    \begin{equation*}
        \begin{array}{lll}
            \alpha_1 = \lambda_1, & \alpha_2 = \lambda_2 x_3, & \alpha_3 = \lambda_3 x_2, \\
            \alpha_4 = \lambda_4 x_1 x_3, & \alpha_5 = \lambda_5, & \alpha_6 = \lambda_6 x_1. 
        \end{array}
    \end{equation*}
    Under the mean-field closure, the means of $X_1,X_2,$ and $X_3$ are governed by the following system of ODEs
    \begin{align}
    \begin{split}
        \frac{\text d \langle X_1 \rangle}{\text d t} &= \lambda_5 - \lambda_6 \langle X_1 \rangle, \\
        \frac{\text d \langle X_2 \rangle}{\text d t} &= \lambda_2 \langle X_3 \rangle - \lambda_3 \langle X_2 \rangle, \\
        \frac{\text d \langle X_3 \rangle}{\text d t} &= \lambda_1 - \lambda_2 \langle X_3 \rangle - \lambda_4 \langle X_1 \rangle \langle X_3 \rangle.
    \end{split}
    \label{eq:chemsig_odes}
    \end{align}
    As demonstrated in~\cite{paulsson_stochastic_2000}, the steady-state behaviour of $\sN$ is determined to a substantial degree by the stochastic fluctuations of $X_3$. This system, therefore, benefits greatly from a hybrid modelling approach, where the low-copy-number $X_1$ and $X_3$ can be modelled discretely. From the CRN $\sN$ we form the M-ARN $\sM$, which has stoichiometry matrix
    \setcounter{MaxMatrixCols}{16}
    \begin{equation*}
        \mathbf M_R = \begin{bmatrix}
        \begin{array}{rrrrrrrrrrrr}
        0 & 0 & 0 & 0 & 0 & 0 & 0 & 0 \\
        0 & 0 & 0 & 0 & 0 & 0 & 0 & 0 \\
        0 & 0 & 0 & 0 & -1 & 0 & 0 & 0 \\
        0 & 0 & 0 & 0 & 0 & 0 & 1 & -1 \\
        0 & 1 & -1 & 0 & 0 & 0 & 0 & 0 \\
        1 & -1 & 0 & -1 & 0 & -1 & 0 & 0
        \end{array} & \rvline & \mathbf M_3
        \end{bmatrix},
    \end{equation*}
    reaction propensities
    \begin{equation*}
        \begin{array}{llll}
            \walpha_{1,1} = \lambda_1, & \walpha_{2,1} = \lambda_2 d_2, & \walpha_{3,1} = \lambda_3 d_3, \\
            \walpha_{4,1} = \lambda_4 d_1 d_2, & \walpha_{4,2} = \lambda_4 d_1 c_2, & \walpha_{4,3} = \lambda_4 c_1 d_2, \\
            \walpha_{5,1} = \lambda_5, & \walpha_{6,1} = \lambda_6 d_1,
        \end{array}
    \end{equation*}
    and the following system of ODEs,
    \begin{align}
        \begin{split}
            \frac{\text d  \langle C_1 \rangle}{\text d t} &= - \lambda_6  \langle C_1 \rangle,\\ 
            \frac{\text d  \langle C_2 \rangle}{\text d t} &= -\lambda_2  \langle C_2 \rangle - \lambda_4  \langle C_1 \rangle  \langle C_2 \rangle,\\
            \frac{\text d  \langle C_3 \rangle}{\text d t} &= \lambda_2  \langle C_2 \rangle - \lambda_3  \langle C_3 \rangle.
        \end{split}
        \label{eq:chemsig_mcmsystem}
    \end{align}
    
\begin{figure}[htbp]
    \centering
    \subfloat[Density plot]{%
        \includegraphics[clip, width=80mm]{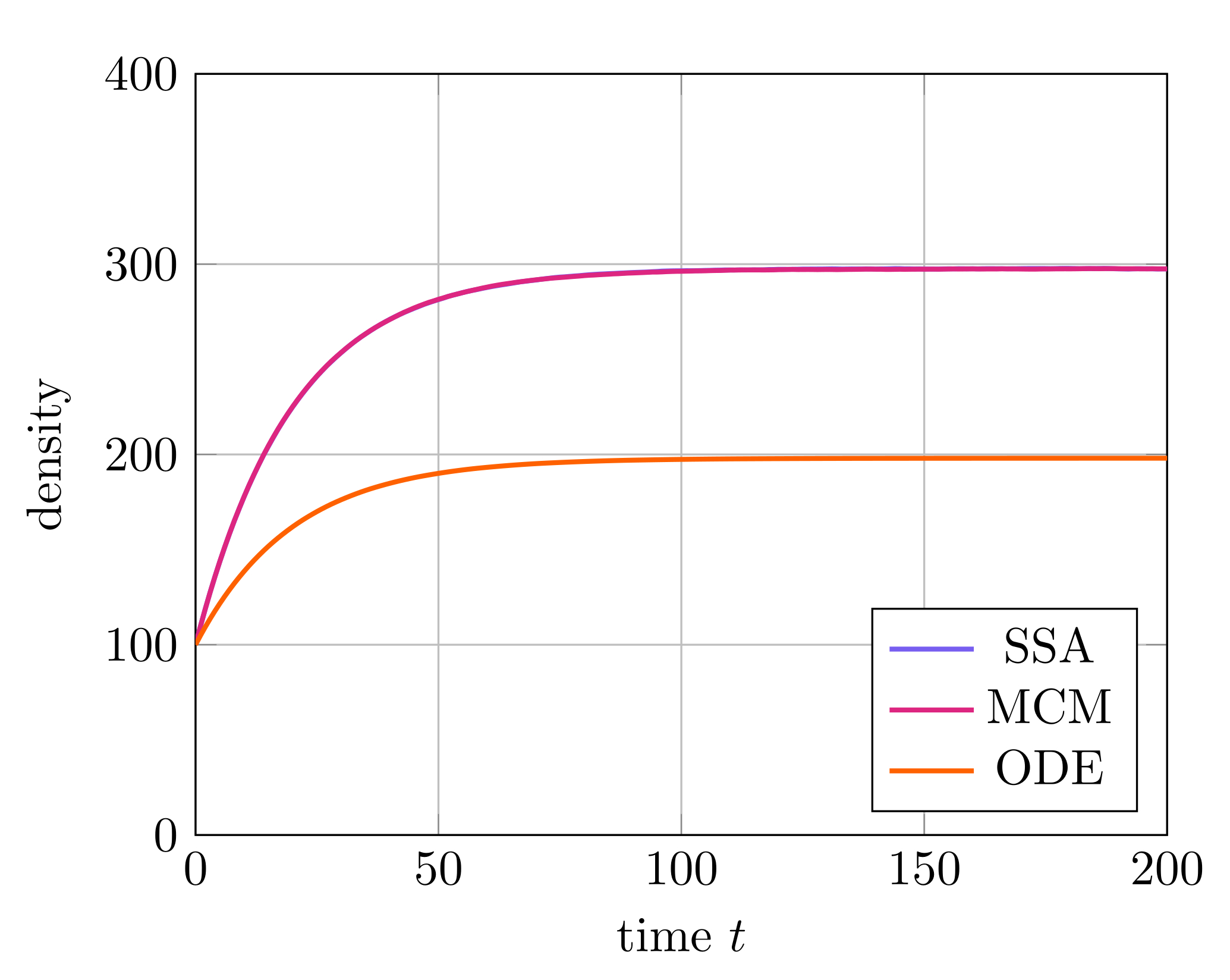}%
    }
    \subfloat[Relative error plot]{%
        \includegraphics[clip, width=80mm]{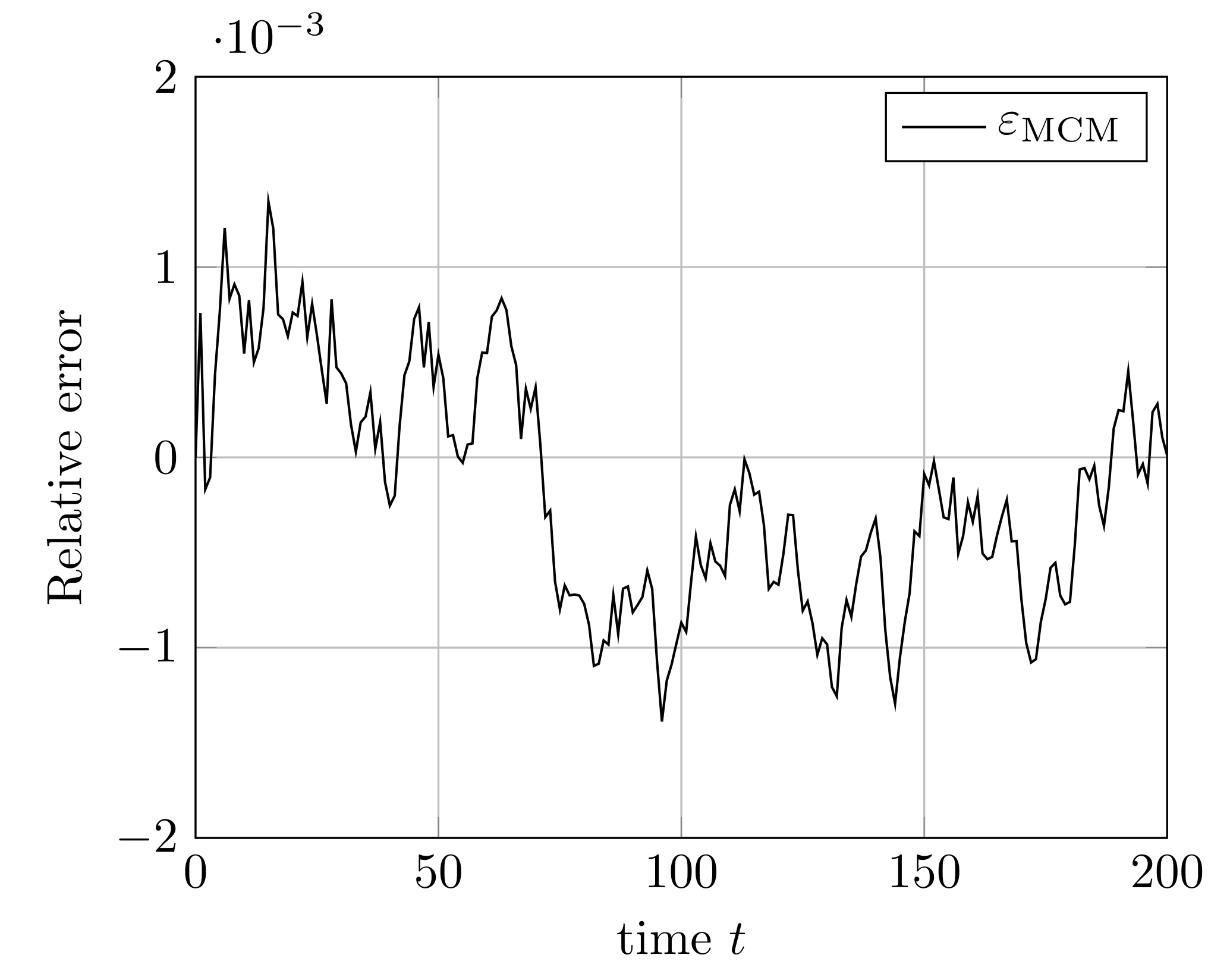}%
    }
    \caption{Results of Test Case 3 (Section~\ref{sec:test-case-3}). (A) Plot of the density of $X_2$ as simulated by the MCM, SSA, and ODE. Note that the density as determined by the SSA is indistinguishable at this scale from the density determined by the MCM - as such the trajectory of the MCM obscures that of the SSA in the plot. (B) Relative error in $X_2$ between the MCM and the SSA. Simulation results averaged over $1.6\cdot 10^4$ repeats.}\label{fig:test-3-3}
    \end{figure}

    To demonstrate the utility of the MCM in this case, we compare the mean densities of $\sN$ as approximated by both the Gillespie SSA and by the mean-field equations~\eqref{eq:chemsig_odes} with the mean density of $\sM$ as approximated by the MCM. For this problem, we wish to simulate the species $X_1$ and $X_3$ purely via the discrete regime and the product species $X_2$ will be permitted to switch regimes dependent on density. The model parameters used for our test case are listed in Table~\ref{table:test-3-params}. We present the results of this test case in Figure~\ref{fig:test-3-3}.
    
    Evidently, the MCM substantially outperforms the mean-field ODEs at approximating the true trajectory of this reaction network. The reason for this is that the MCM guarantees the simulation of the $X_3$ species exclusively via the discrete regime by setting the relevant regime conversion threshold values to infinity. As such, the method retains information of the stochastic fluctuations in $X_3$ where the system of mean-field ODEs does not. We further note the lack of bias in the error of the MCM.

\section{Discussion} %!%--4 Discussion--%%%%%%%%%%%%%%%%%%%%%%%%%%%%%%%%%%%%%%%%%%%%%%%%%%%%%%%%%%%%%%%%%%%%%%%%%%%%%%%%
  In this work we introduced a novel hybrid method for simulating well-mixed chemical reaction networks. This method couples a system of ODEs with a Markov process representation of a chemical reaction network by constructing a so-called \textit{augmented reaction network} that combines both representations. The continuous and discrete components of the augmented network can be simulated simultaneously using different techniques to maximise computational efficiency and minimise the loss of accuracy resultant from taking continuum approximations. We demonstrated the accuracy of the method in three separate test problems of increasing complexity, evidencing in the final test case a substantial improvement in accuracy using our method versus the standard continuum approximation technique.

  While our method demonstrates substantially better accuracy versus the continuum-only models in the test cases we present, its advantage versus a traditional SSA is, in general, dependent on network structure. Specifically, in systems where the majority of computation time (when simulated via a SSA) is spent on the simulation of low copy-number species interacting with high copy-number species via bimolecular reaction channels, there is little computational benefit to our approach. The reason for this is that such reactions are (assuming each species is below and above the transition thresholds, respectively) necessarily simulated using the SSA, and therefore may impart no computational benefit in the MCM versus the SSA alone. In cases where both reactant species in a bimolecular reaction are of sufficiently high concentration to be above their respective transition thresholds, it may be the case that the MCM yields similar accuracy to that of a continuum-based approach. Nevertheless, in neither case is there reason to expect \textit{a priori} that use of the MCM is necessarily disadvantageous. With these caveats in mind, there are clear instances where the MCM may be suitable to use over traditional methods. In loosely-coupled networks where the majority of interactions are of first-order (networks of this type frequently arise when modelling cellular populations \cite{simpson_migration_2010,simpson_cell_2010,yates_multi-stage_2017,kynaston_equivalence_2022}), the MCM demonstrates a clear computational advantage.

  There are several ways in which the MCM might be extended to accommodate a wider variety of problems and to increase its computational efficiency. The first and most obvious direction is to extend its dimensionality; for example, to a spatial setting. The MCM, being an effective simulation technique for well-mixed reaction networks, might be extended to a spatial reaction-diffusion setting in several ways. Under a mesoscopic modelling regime (see e.g.,~\cite{smith_spatially_2018}), where indivdual system components are collected into well-mixed spatial `bins' of fixed size, the MCM could be used to simulate reactions by treating individuals in each bin as distinct species that do not interact with neighbouring bins. In this framework, diffusive jumps between bins are simply reactions that convert individuals in one `bin species' to another. A spatial model consisting of binned particles and ordinary differential equations associated with each bin is thereby easily teated via the MCM. Nevertheless, this representation of a reaction-diffusion process is limited - for spatial domains with many bins, simulating large systems of (potentially) non-linear ODEs may be prohibitively expensive. A more sensible choice would be to represent the continuous approximation as a system of partial differential equations on an explicitly spatial domain; indeed, contemporary spatially-extended hybrid methods that couple continuous and mesoscopic regimes generally use this representation~\cite{smith_spatially_2018}. In this case, the matter of coupling the stochastic and diffusive reactions in each bin is not so straightforward, requiring numerical integration of the partial differential equation over relevant spatial regions. Extending the MCM in this manner to a spatially-extended mesoscopic-to-continuous hybrid method will form the basis of an upcoming investigation.

  The MCM may also be extended to incorporate additional dimensionality along non-spatial lines. An important class of demographic and biological models are those with size- or age-structure, or a combination thereof. These model systems of interacting individuals (either eukaryotic or prokaryotic cells) undergoing some variant of the classical cell-cycle~\cite{schafer_cell_1998}, and for which an individual's size (or age) is an important contributor to overall population dynamics. These systems are often modelled as either discrete-state stochastic processes~\cite{stukalin_age-dependent_2013,greenman_kinetic_2016,chou_hierarchical_2016,kynaston_equivalence_2022} or as continuous partial differential equations via the McKendrick-von Foerster equation~\cite{trucco_mathematical_1965,rossini_novel_2021}. Despite their ubiguity, to the best of the authours' knowledge there exist no hybrid simulation techniques that can accommodate, without modification, size- or age-structure. Depending on the specific functional form of any size- or age-mediated reactions, a method of `spatial' numerical integration over intervals of age or size similar to that proposed for spatial extension may prove fruitful for coupling these two modelling regimes.

  An important area of research in numerical methods in general is the development of so-called `adaptive' methods. These are methods for which certain numerical parameters can be changed mid-way through a simulation run to adapt to situations that might otherwise prove numerically challenging or computationally infeasable. The prototypical example of this is in adaptive time-stepping methods for solving systems of ordinary differential equations, wherein the usual fixed time step of a numerical solver is replaced with a variable time step that is recalculated at each update step to ensure stability even when the derivates of the system undergo large variations \cite{alexander-solving-1990}. As noted in our description of the MCM, one could apply such an adaptive time-stepping method for computing approximations to the continuum regime description without needing to modify the algorithm. Nevertheless, it is not difficult to concieve of a more specialised form of time-stepping that would take into consideration more than just the mass in the continuous regime and instead consider both the discrete mass and the calculated propensity functions at the time of an update step. For example, large numbers of individuals either entering or leaving the continuous regime may, by affecting the gradient of the continuum approximation, inject undesirable numerical instabilities into the MCM in extreme cases - something that traditional adaptive time-stepping methods are not designed to handle. Adaptivity in terms of time-stepping is not the only potential improvement, however. Presently, the MCM requires that the threshold values for the regime conversion reactions are set and fixed \textit{a priori}. One can envisage modifications to the MCM where the conversion thresholds vary in response to changes in density, computational cost, rates of density change, and the stochasticitiy present in the system at any given time.

  Finally, the method may be extended to incorporate reactions of arbitrary molecular order. While any reactions of molecular order of at least three can be decomposed into sequences of reactions of molecular order of at most two, these decompositions can be difficult to compute in practice. We conjecture that the same techniques used to demonstrate equivalence between the CRN and its associated ARN apply to higher-order reactions; however, proving this in generality is likely to be cumbersome. Further, one needs $2^d$ ODEs to satisfy the coupling requirements~\ref{hypothesis-1},~\ref{hypothesis-2} for a reaction of order $d$ which, while not necessarily impacting computation time, may quickly become impractical to implement for large networks.
  
  To summarise, our method provides a novel and computationally efficient technique for simulating well-mixed chemical reactions networks using a hybrid discrete/continuous methodology. Unlike similar existing methods, ours does not depend on the system of interest possessing certain properties; i.e., a particular decomposability of reactions or species into `fast' and `slow' categories. Further, it represents a promising coupling mechanism between the mesoscopic and macroscopic regimes that may permit for the development of new spatially-extended hybrid techniques that have a particular intrinisic adaptivity; namely, the ability to simulate spatial density distributions with significant and dynamic heterogeneity. 

\footnotesize
\bibliographystyle{ieeetr}
\bibliography{references}

\newpage

\section*{List of Figures} %!%--List of Figures--%%%%%%%%%%%%%%%%%%%%%%%%%%%%%%%%%%%%%%%%%%%%%%%%%%%%%%%%%%%%%%%%%%%%%%%

\section*{List of Tables} %!%--List of Tables--%%%%%%%%%%%%%%%%%%%%%%%%%%%%%%%%%%%%%%%%%%%%%%%%%%%%%%%%%%%%%%%%%%%%%%%%%
  \begin{algorithm}
    \caption{The mass-conversion method}
    \label{alg:mcm}
    \begin{algorithmic}[1]
      \State{Specify initial conditions $\bm Y(0) = \bm{C}(0) \oplus \bm{D}(0)$ and $t_0$, final time $t_f$, and ODE update step size $\Delta t$.}
      \State{Set $t_d = \Delta t$}
      \While{$t < t_f$}
          \State{Calculate the value of each reaction propensity function $\widetilde{\alpha}_{r,k}(\bm Y(t))$}
          \State{Calculate the value of each conversion reaction propensity function $\widetilde{\alpha}_{f,i}(\bm Y(t))$ and $\widetilde{\alpha}_{b,i}(\bm Y(t))$}
          \State{Calculate the sum of all propensity functions at time $t$ \[\alpha_0 = \sum_{r\in\sR} \sum_{j=1}^{2^d}{\widetilde\alpha}_{r,j} + \sum_{i=1}^K \left(\widetilde{\alpha}_{f,i} + \widetilde{\alpha}_{b,i}\right)\]}
          \State{Sample uniformly at random a number $u$ from the interval $[0,1]$}
          \State{Determine the time until the next stochastic event \[\tau = \frac{1}{\alpha_0} \ln \left(\frac{1}{u}\right)\]}
          \If{$t + \tau < t_d$} \Comment{The next stochastic event occurs}
              \State{Determine which event occurs by finding $j$ such that \[ \sum_{p=1}^j \widetilde{\alpha}_p(t) < r\alpha_0 < \sum_{p=1}^{j+1} \widetilde{\alpha}_p(t). \]}
              \If{The firing of reaction $j$ would result in $C_k < 0$ for some $k$}
              \If{$C_k<u$ for $u \sim U(0,1)$}
              \State{The reaction is not executed.}
              \Else{}
              \State{Update the state via $\bm Y(t+\tau) = \bm Y(t) + \nu_p$ and set $C_k = 0$.}
              \EndIf{}
              \Else{}
              \State{Update the state via $\bm Y(t+\tau) = \bm Y(t) + \nu_p$.}
              \EndIf{}
              \State{Set $t = t + \tau$.}
          \Else{} \Comment{The next ODE update occurs}
              \State{Perform an ODE update step to calculate $\bm{c}(t_d)$}
              \State{Set $t = t_d$}
              \State{Set $t_d = t + \Delta t$}
          \EndIf{}
      \EndWhile{}
    \end{algorithmic}
  \end{algorithm}

  \newcolumntype{R}{>{\raggedleft\arraybackslash}X}
  \newcolumntype{C}{>{\centering\arraybackslash}X}
  \newcolumntype{L}{>{\raggedright\arraybackslash}X}
  \renewcommand{\arraystretch}{1.3}

  \begin{table}
      \centering 
      \begin{tabularx}{\linewidth}{l|Cr}
          \toprule[1pt]\midrule[0.3pt]
          & Variable & Value \\
          \midrule
          \multirow{2}{*}{Initial conditions} & $C$ & $1.0\cdot 10^3$ \\
          & $D$ & $0$ \\
          \midrule
          \multirow{3}{*}{Reaction rates} & $\lambda_1$ & $1.0\cdot 10^0$ \\
          & $\lambda_2$ & $2.0\cdot 10^2$ \\
          & $\gamma$ & $1.0\cdot 10^1$ \\
          \midrule
          \multirow{1}{*}{Threshold values} & $T_1^f = T_1^b$ & $6.5\cdot 10^2$ \\ 
          \midrule
          \multirow{2}{*}{Simulation parameters} & $\Delta t$ & $1.0\cdot 10^{-4}$\\
          & $t_f$ & $8.0\cdot 10^{0}$\\
          \midrule
          \multirow{2}{*}{Other} & $I_1$ & $[4,\infty)$ \\
          & $I_2$ & $[0,4)$ \\
          \midrule[0.3pt]\bottomrule[1pt]
      \end{tabularx}
      \caption{Initial and parameter values for Test Problem 1.}\label{table:test-1-params}
  \end{table}

  \begin{table}
    \centering 
    \begin{tabularx}{\linewidth}{l|Cr}
        \toprule[1pt]\midrule[0.3pt]
        & Variable & Value \\
        \midrule
        \multirow{2}{*}{Initial conditions} & $C$ & $0.0\cdot 10^0$ \\
        & $D$ & $6.0\cdot 10^1$ \\
        \midrule
        \multirow{3}{*}{Reaction rates} & $\lambda_1$ & $1.0\cdot 10^{-3}$ \\
        & $\lambda_2$ & $6.0\cdot 10^{-1}$ \\
        & $\gamma$ & $1.0\cdot 10^0$ \\
        \midrule
        \multirow{2}{*}{Threshold values} & $T_1^f$ & $6.5\cdot 10^2$ \\ 
        & $T_1^b$ & $7.0\cdot 10^2$ \\
        \midrule
        \multirow{2}{*}{Simulation parameters} & $\Delta t$ & $1.0\cdot 10^{-2}$\\
        & $t_f$ & $8.0\cdot 10^{0}$\\
        \midrule
        \multirow{1}{*}{Other} & $I$ & $[0,20) \cup [40,60)$ \\
        \midrule[0.3pt]\bottomrule[1pt]
    \end{tabularx}
    \caption{Initial and parameter values for Test Problem 2.}\label{table:test-2-params}
  \end{table}

  \begin{table}
    \centering 
    \begin{tabularx}{\linewidth}{l|Cr}
        \toprule[1pt]\midrule[0.3pt]
        & Variable & Value \\
        \midrule
        \multirow{3}{*}{Initial conditions}
        & $D_1$ & $1.0\cdot 10^1$ \\
        & $D_2$ & $1.0\cdot 10^2$ \\
        & $D_3, C_1, C_2, C_3$ & 0 \\
        \midrule
        \multirow{7}{*}{Reaction rates} 
        & $\lambda_1$ & $1.0 \cdot 10^2$ \\ 
        & $\lambda_2$ & $1.0\cdot 10^3$ \\
        & $\lambda_3$ & $1.0\cdot 10^{-2}$ \\ 
        & $\lambda_4$ & $9.9\cdot 10^{3}$ \\ 
        & $\lambda_5$ & $5.0\cdot 10^2$ \\
        & $\lambda_6$ & $1.0\cdot 10^2$ \\
        & $\gamma$ & $1.0\cdot 10^0$ \\
        \midrule
        \multirow{3}{*}{Threshold values}
        & $T_{f,1}, T_{b,1}, T_{f,3}, T_{b,3}$ & $\infty$ \\ 
        & $T_{f,2}$ & $2.0\cdot 10^2$ \\ 
        & $T_{b,2}$ & $2.5\cdot 10^2$ \\
        \midrule
        \multirow{2}{*}{Simulation parameters} & $\Delta t$ & $1.0\cdot 10^{-1}$\\
        & $t_f$ & $2.0\cdot 10^{2}$\\
        \midrule[0.3pt]\bottomrule[1pt]
    \end{tabularx}
    \caption{Initial and parameter values for Test Problem 3.}\label{table:test-3-params}
  \end{table}

  % \begin{table}
  %     \centering
  %     \begin{tabularx}{\linewidth}{LC|CR}
  %     \toprule[1pt]\midrule[0.3pt]
  %     Variable & Initial Value & Rate & Value \\ \midrule
  %     $X_1$ & 100 & $k_1$ & $1.0 \cdot 10^2$ \\
  %     $X_2$ & 0 & $k_2$ & $1.0\cdot 10^3$ \\
  %     $X_3$ & 100 & $k_3$ & $1.0\cdot 10^{-2}$ \\
  %     &  & $k_4$ & $9.9\cdot 10^{3}$ \\
  %     &  & $k_5$ & $5.0\cdot 10^{2}$ \\
  %     &  & $k_6$ & $1.0\cdot 10^2$ \\
  %     &  & $N_{13}$ & $2.0 \cdot 10^2$ \\
  %     &  & $N_{23}$ & $2.5 \cdot 10^2$ \\
  %     &  & $r_1, r_2, r_3$ & $1.0 \cdot 10^4$ \\
  %     &  & $N_{11}, N_{21}, N_{12}, N_{22}$ & $\infty$ \\ \midrule[0.3pt]\bottomrule[1pt]
  % \end{tabularx}
  % \caption{Initial and parameter values for Test Problem 3.}\label{table:test-3-params}
  % \end{table}

\end{document}